\documentclass [10pt,compsocconf]{IEEEtran}
\usepackage{algorithm,amsbsy,amsmath,amssymb,epsfig,bbm,mathrsfs,multirow,amsthm}
\usepackage{algpseudocode}
\usepackage{subcaption,float}
\usepackage[hidelinks]{hyperref}
\newtheorem{proposition}{Proposition} 
\newtheorem{assumption} {Assumption}

\renewcommand{\baselinestretch}{0.95}
\begin{document}
\title{Reinforcement Learning Approach for RF-Powered Cognitive Radio Network with Ambient Backscatter}
\author{\IEEEauthorblockN{Nguyen Van Huynh$^1$, Dinh Thai Hoang$^1$, Diep N. Nguyen$^1$, Eryk Dutkiewicz$^1$, Dusit Niyato$^2$, and Ping Wang$^2$ \\}
	$^1$ School of Electrical and Data Engineering, University of Technology Sydney, Australia\\
	$^2$ School of Computer Science and Engineering, Nanyang Technological University, Singapore 	\vspace{-5mm}}	

\maketitle
\thispagestyle{empty}
\begin{abstract}
For an RF-powered cognitive radio network with ambient backscattering capability, while the primary channel is busy, the RF-powered secondary user (RSU) can either backscatter the primary signal to transmit its own data or harvest energy from the primary signal (and store in its battery). The harvested energy then can be used to transmit data when the primary channel becomes idle. To maximize the throughput for the secondary system, it is critical for the RSU to decide when to backscatter and when to harvest energy. This optimal decision has to account for the dynamics of the primary channel, energy storage capability, and data to be sent. To tackle that problem, we propose a Markov decision process (MDP)-based framework to optimize RSU's decisions based on its current states, e.g., energy, data as well as the primary channel state. As the state information may not be readily available at the RSU, we then design a low-complexity online reinforcement learning algorithm that guides the RSU to find the optimal solution without requiring prior- and complete-information from the environment. The extensive simulation results then clearly show that the proposed solution achieves higher throughputs, i.e., up to 50\%, than that of conventional methods.
\end{abstract}

{\it Keywords-} Ambient backscatter, RF energy harvesting, cognitive radios, MDP, reinforcement learning. 
 
\section{Introduction}
\label{sec:Introduction}

Radio frequency (RF) powered cognitive radio networks (CRNs) have been seen as an emerging solution to address both the radio spectrum shortage and the energy limitation for low-power secondary systems (e.g., in industrial IoT applications). In an RF-powered CRN, while the primary transmitter, e.g., the base station, broadcasts signals to its receivers, the secondary transmitter (ST) can harvest energy from such signals through RF energy harvesting techniques. The harvested energy is then stored in the battery of the ST and used to transmit its own data to the secondary receiver (SR) when the primary channel becomes idle, i.e., the base station ceases broadcasting. In this way, the secondary system can operate with minimal human intervention and without causing any interference to the primary system. As a result, there are paramount applications of RF-powered CRNs in practice such as low-energy sensor and IoT networks~\cite{Niyato2016WirelessBook}. However, in an RF-powered CRN, the performance of secondary system heavily depends on the activities of the primary channel that controls both energy and radio frequency of STs. In particular, when the primary channel is usually busy, i.e., the base station broadcasts signals most of the time, the ST has very limited opportunities to transmit data, resulting in a low throughput. This problem can be tackled by recent advances in ambient backscattering. 

Ambient backscatter communication (ABC) allows wireless devices to communicate by modulating and reflecting the surrounding ambient RF signals~\cite{LiuAmbient2013}. The ABC technology bears close resemblance with radio frequency identification (RFID), but while RFID requires transmissions from a dedicated carrier emitter, ABC can modulate surrounding ambient signals transmitted by existing wireless systems. Hence, ABC systems can share spectrum with exiting systems and achieve better spectral efficiency than that of RFID systems. Furthermore, ABC devices are relatively simple and consume much less power than active transmitters, and thus ABC allows ultra-low-power operation with low cost implementation~\cite{Huynh2017Survey}. As a result, ABC technology has been receiving significant attention recently, and it was listed as one of the 10 breakthrough technologies in 2016 by MIT Technology Review~\cite{MIT_Tech_Review}. For RF-powered CRNs that employ ABC, while the primary channel is mostly busy, instead of spending whole time to harvest energy, the ST can use a fraction of time to transmit data by modulating and backscattering the received signals through ABC technique. Thus, ABC enables secondary systems to simultaneously optimize the spectrum usage and energy harvesting to maximize their performance. 

There were some research works in the literature studying solutions to integrate ABC into RF-powered CRNs. In~\cite{LiuAmbient2013}, the authors introduced a circuit diagram together with a prototype for an ambient backscattering device with RF energy harvesting capability, i.e., ST.  This device includes three main components, i.e., an antenna, an energy harvesting circuit, and a controller. The prototype device can achieve information rates of 1 $Kbps$ over the distances of 2.5 feet. The authors in~\cite{Parks2014Turbocharging} then extended~\cite{LiuAmbient2013} by introducing a novel coding scheme to improve the backscatter transmission rate as well as the communication range. In this technique, each data bit is represented by one symbol, and each symbol in turn is represented by a predefined chip sequence. Through experiments, the authors showed that the backscatter transmission rate and the communication range can be extended up to 1 Mbps and 20 meters, respectively.

Some other solutions were also proposed to improve the performance for secondary systems. In~\cite{Kim2017Hybrid}, a hybrid backscatter communications for RF-powered CRNs was introduced in order to improve transmission range and rate for the secondary system. In the network under consideration, the ST can flexibly select between an ambient RF source or a dedicated RF source to support its transmissions based on its location, i.e., indoor-zone or outdoor-zone. Then, an energy trade-off problem is formulated to maximize the throughput for the hybrid backscatter communications. In~\cite{Hoang2017Ambient}, the time trade-off problem between the harvest-then-transmit and backscatter processes for an RF-powered backscatter CRN was studied. The numerical results demonstrate that the integration of ambient backscatter technique into RF-powered CRNs always achieves the higher transmission rate than that of using either the ambient backscatter communication or the harvest-then-transmit scheme alone.

For RF-powered CRN with ABC, the optimal decision of ST on when to backscatter, when to harvest, when to transmit has to account for the dynamics of primary channel state, energy/battery status, data to be transmitted. Unfortunately, these dynamics are either not readily available at a ST or difficult to be predicted. In this paper, we develop a low-complexity online reinforcement learning algorithm to deal with these dynamics of the environment and aim to maximize the ST's throughput. Specifically, we first formulate the optimal decision problem for the ST as a Markov decision process. We then develop an online learning algorithm which enables the ST to find the optimal policy through ``learning'' from its interactions with the environment. Through simulation results, we demonstrate that our proposed learning algorithm achieves the best performance compared to existing methods and close to that of the optimal solution achieved when all environment information is known in advance.

\section{System Model}
\label{sec:sysmodel}

\begin{figure}[!]
	\captionsetup{singlelinecheck=off}
	\centering
	\includegraphics[scale=0.17]{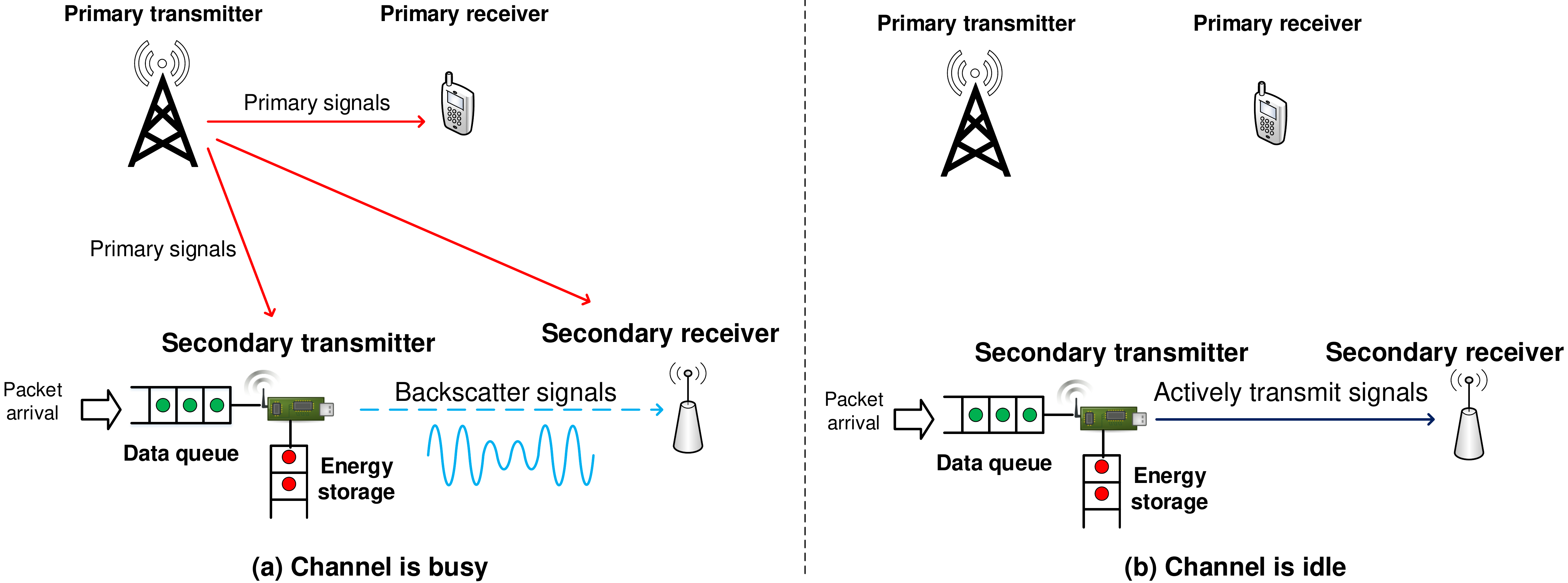}
	\caption{System model.}
	\label{Fig.system_model}
\end{figure}

Consider a primary system and a secondary system coexisting in an area as shown in Fig.~\ref{Fig.system_model}. The secondary system consists of a secondary transmitter (ST) which wants to transmit data to its secondary receiver (SR). The ST is equipped with RF energy harvesting and ambient backscatter circuits. While the primary channel is busy, the ABC allows the ST to either harvest energy from the primary signals (to store in its energy storage) or backscatter the signals to transmit data as shown in Fig.~\ref{Fig.system_model}(a). In contrast, while the channel is idle, i.e., Fig.~\ref{Fig.system_model}(b), the ST can actively transmit data to its SR by using the energy in its energy storage. Let $E$ and $D$ be the maximum energy storage capacity and maximum data queue size of the ST, respectively. In each time slot, a packet arriving at the data queue with probability $\alpha$. The probability of the primary channel being idle is denoted by $\eta$. When the channel is busy and the ST performs backscattering, i.e., backscatter policy, the ST can transmit $d_\mathrm{b}$ data units successfully with probability $\beta$. However, if the ST chooses to harvest energy in the busy period, it can harvest $e_\mathrm{h}$ units of energy successfully with probability $\gamma$. When the channel becomes idle, the ST can use $e_\mathrm{t}$ units of energy to successfully transmit $d_\mathrm{t}$ data units to its receiver with probability $\sigma$. This process is also known as harvest-then-transmit (HTT) mode~\cite{park2013}. Note that our proposed system model can be straightforwardly extended to multiple STs that operate on different primary channels to avoid collision.
In the proposed system, two successive working periods of the PT, i.e., idle and busy, are taken into account. As mentioned, the ST can choose to harvest energy or backscatter data in busy periods, and actively transmit data in idle periods. This leads to a trade-off problem among data backscattering, energy harvesting, and data transmitting time to achieve the optimal network throughput. Intuitively, based on its current state, i.e., the data queue state, the energy storage state, and the primary channel state, the ST needs to make a decision to transmit data, harvest energy, backscatter data, or stay idle. However, in practice, the environment parameters, e.g., channel idle probability and successful data transmission probability, may not be available in advance. Therefore, in the following, we introduce an online learning algorithm that can help the ST make the optimal decisions without requiring the complete environment parameters.

\section{Problem Formulation}
\label{sec:LA}

\subsection{MDP Description}
We define the state space of the ST as follows:
\begin{equation}
\begin{split}
\mathcal{S} =	\Big\{ ({\mathcal{C}}, {\mathcal{D}}, {\mathcal{E}} ); {\mathcal{C}} \in \{0,1\}, {\mathcal{D}} \in \{0,\ldots,d, \ldots, D\},\\
{\mathcal{E}} \in \{0,\ldots,e, \ldots, E \} \Big\},
\end{split}
\end{equation}
where $c \in \mathcal{C}$ represents the state of the primary channel, i.e., $c = 1$ when the primary channel is busy and $c = 0$ otherwise, $d \in {\mathcal{D}}$ and $e \in {\mathcal{E}}$ represent the number of data units in the data queue and the energy units in the energy storage of the ST, respectively. Then, we define the state of the ST as a 3-tuple $s = (c,d,e) \in \mathcal{S}$, where $c$, $d$ and $e$ are the channel state, the data state, and the energy state, respectively. As mentioned, the ST can choose one of four actions, i.e., harvest energy, transmit data, backscatter data, or stay idle, to perform. Therefore, we define the action space of the ST as follows:
\begin{equation}
\mathcal{A}	\triangleq \{a:a \in \{1,\ldots, 4\} \}	,
\end{equation}
where
\begin{equation}
a 	=	\left\{	\begin{array}{ll}
1,	&	\mbox{when the ST stays idle},	\\
2,	&	\mbox{when the ST transmits data},	\\
3,	&	\mbox{when the ST harvests energy},	\\
4,	&	\mbox{when the ST backscatters data}.
\end{array}	\right.
\end{equation}
Moreover, when the ST is in state $s$, its action space is denoted by $\mathcal{A}_s$. Note that $\mathcal{A}_s$ consists of \emph{feasible} actions that do not lead a transition to an unreachable state. Therefore, $\mathcal{A}_s$ can be defined as follows:
\begin{equation}
\mathcal{A}_s	=	\left\{	\begin{array}{ll}
\{1\},	&	\mbox{if $c=0$ and $d<d_\mathrm{t}$}\\
& \mbox{OR $c=0$ and $e < e_\mathrm{t}$}\\
&\mbox{OR $c=1$, $e=E$ and $d<d_\mathrm{b}$},\\
\{1,2\},	&	\mbox{if $c=0$, $d \geq d_\mathrm{t}$ and $e \geq e_\mathrm{t}$},	\\
\{3\},	&	\mbox{if $c=1$, $d < d_\mathrm{b}$ and $e<E$},	\\
\{4\},	&	\mbox{if $c=1$, $d \geq d_\mathrm{b}$ and $e=E$},	\\
\{3,4\},	&	\mbox{if $c=1$, $d \geq d_\mathrm{b}$ and $e<E$}.
\end{array}	\right.
\end{equation}
The first condition refers to the case when the primary channel is idle and there is not enough data, e.g., no data, or insufficient energy for active transmission. This condition also applies to a special case when the energy storage is full, the primary channel is busy, and the number of data units in the no data for backscattering. Thus, the ST can only select to stay idle, i.e., $a=1$. The second condition corresponds to the case in which the primary channel is idle and there are data and sufficient energy to perform active transmission. When the primary channel is busy, if there is not enough data, e.g., no data, for backscattering, and the energy storage is not full, the ST will choose to harvest energy, i.e., the third condition. Otherwise, if there is data to backscatter, the ST can choose to backscatter data or harvest energy if the energy storage is not full, i.e., the fourth and fifth conditions.

When the ST successfully transmits or backscatters data to its receiver, it will receive an immediate reward, i.e., throughput ${\mathcal{T}}$, denoted as follows:
\begin{equation}
{\mathcal{T}} (s,a)	=	\left\{	\begin{array}{ll}
\sigma d_\mathrm{t},	&	(a=2),	\\
\beta d_\mathrm{b},	&	(a=4),	\\
0	,						&	\mbox{otherwise}	.
\end{array}	\right.
\end{equation}

When all environment parameters, e.g., channel idle probability and successful data transmission, are known, we can derive the transition probability matrix for the MDP and use conventional algorithms~\cite{Puterman1995}, e.g., value iteration algorithm, to obtain the optimal policy for the ST. However, in practice, some environment parameters may not be available in advice. As a result, we are unable to derive the transition probability matrix for the MDP. In the following, we propose the reinforcement online learning algorithm to resolve this issue. The optimal policy obtained by the MDP using value iteration algorithm will be used as a benchmark to evaluate the performance of the proposed solution.
\subsection{Parameterization for the MDP}
We consider a randomized parameterized policy~\cite{Marbach2001} with \textit{softmax} action selection rules~\cite{Sutton1998Reinforcement} to find decisions for the ST.
With the randomized parameterized policy, the ST will choose action $a$ at state $s$ with the normalized probability as follows:
\begin{equation}
\chi_{\Theta}(s,a) = \frac{\exp\big(\theta_{ s,a }\big)} {\sum_{a' \in \mathcal{A}}\exp\big(\theta_{ s,a'  }\big)},
\label{eq:randomized_parameterized_function}
\end{equation}
where $\Theta = \left[	\begin{array}{ccc}	\cdots	&	\theta_{s, a}	&	\cdots	\end{array}	\right]^\top$ is the parameter vector of the learning algorithm. By interacting with the environment, the algorithm will update this parameter vector iteratively. Furthermore, $\chi_{\Theta}(s,a)$ must not be negative and meets the following constraint: 
\begin{equation}
\label{eq:condition_1}
\sum_{a \in \mathcal{A}} \chi_{\Theta}(s,a) = 1	. 
\end{equation} 
The parameterized immediate throughput function of the ST is then as follows:
\begin{equation}
\label{eq:average_throguhput_defination}
{\mathcal{T}}_{\Theta} ( s )	= \sum_{a \in \mathcal{A}} \chi_{\Theta}(s, a){\mathcal{T}}( s, a ) ,
\end{equation}	
where $\mathcal{T}( s, a )$ denotes the immediate throughput. Similarly, the parameterized transition probability function can also be derived as follows:
\begin{equation}
P_{\Theta}({s,s'	})= \sum_{a \in \mathcal{A}} \chi_{\Theta}(s, a)	P_{s,s'}(  a ), \quad \forall s, s' \in \mathcal{S}, 
\end{equation}
where $P_{s,s'}( a )$ is the transition probability from state $s$ to state $s'$ when action $a$ is taken. After that, the average throughput of the ST can be parameterized as follows:
\begin{equation}
\label{eq:throughput}
\xi(\Theta) = \lim_{t\rightarrow \infty} \frac {1}{t} \mathbb{E}_{\Theta} \Big[ \sum_{k=0}^{t} {\mathcal{T}}_{\Theta} ( s_k )\Big] ,
\end{equation}
where $s_k$ is the state of the ST at time step $k$. $\mathbb{E}_{\Theta}[ \cdot ]$ is the expectation of the throughput. Then, we make following assumptions:
\begin{assumption}
	\label{recurrent_state}
	There exists a recurrent state $s^{*}$ which is visited by the online learning algorithm for each of the Markov chain, and this Markov chain needs to be aperiodic.
\end{assumption}

Assumption~\ref{recurrent_state} ensures that the considered system has a Markov property. Additionally, we have the following balance equations:
\begin{equation}
\begin{aligned}
\label{eq: balance equation}
&\sum_{s \in \mathcal{S}} \pi_{\Theta}({s})=1 \mbox{ and} \quad \sum_{  s \in \mathcal{S} } \pi_{\Theta}({s}) P_{\Theta}({s,s'}) = \pi_{\Theta}({s'}),
\forall s' \in \mathcal{S},
\end{aligned}
\end{equation}
where $\pi_{\Theta}({s})$ is the steady-state probability of state $s$ under the parameter vector $\Theta$. With (\ref{eq:throughput}) and (\ref{eq: balance equation}), we can express the parameterized average throughput as follows:
\begin{equation}
\label{eq:average throughput}
\xi(\Theta) = \sum_{s	\in		\mathcal{S}} \pi_{\Theta}({s}) {\mathcal{T}}_{\Theta} ( s )	.
\end{equation}
We aim to maximize $\xi(\Theta)$ given the parameter vector $\Theta$.
\subsection{Policy Gradient Method}
We define the differential throughput $d(s,\Theta)$ at state $s$ as follows:
\begin{equation}
d(s, \Theta) = \mathbb{E}_{\Theta} \left[ \sum_{k=0}^{T-1} \left( 	{\mathcal{T}}_{\Theta} ( s_k ) - \xi(\Theta)	\right) | s_{0} = s \right],
\end{equation}
where $T=\min\{k>0 | s_k = s^{*}\}$ is the first future time that the online learning algorithm visits the recurrent state $s^{*}$. Then, with the differential throughput $d(s,\Theta)$, the gradient of the average throughput $\xi(\Theta)$ can be easily derived as stated in Proposition~\ref{prop_policy_gradient}.
\begin{proposition}
	\label{prop_policy_gradient}
	Under Assumption~\ref{recurrent_state} and Assumption~\ref{derivatives}, we have
	\begin{equation}
	\nabla \xi(\Theta) = \sum_{s \in \mathcal{S}} \pi_{\Theta}(s) \Big(\nabla {\mathcal{T}}_{\Theta} ( s ) + \sum_{s' \in \mathcal{S}} \nabla P_{\Theta}(s,s') d(s',\Theta)  \Big) .
	\end{equation}
\end{proposition}
The proof of Proposition~\ref{prop_policy_gradient} can be found in~\cite{Marbach2001}. In addition, we make an assumption as follows:
\begin{assumption}
	\label{derivatives}
	For every state $s, s' \in \mathcal{S}$, the immediate throughput function ${\mathcal{T}}_{\Theta} ( s )$ and the transition probability function $P_{\Theta}({s,s'})$ satisfy the following conditions: (1) twice differentiable and (2) the first and second derivatives are bounded.
\end{assumption}
Assumption~\ref{derivatives} ensures that the average throughput is well defined for every $\Theta$ and does not depend on the initial state.
\subsection{Idealized Gradient Algorithm}
As stated in~\cite{Bertsekas1995}, the idealized gradient algorithm is formulated through Proposition~\ref{prop_policy_gradient} as follows:
\begin{equation}
\label{idealized_algorithm_theta}
\Theta_{k+1} = \Theta_{k} + \rho_{k} \nabla \xi(\Theta_{k}),
\end{equation}
where $\rho_{k}$ is a step size satisfied Assumption~\ref{ass:step size}.
\begin{assumption}
	\label{ass:step size}
	The step size $\rho_{k}$ is nonnegative, deterministic, and satisfies
	\begin{equation}
	\sum_{k=1}^{\infty}\rho_{k} = \infty, \mbox{ and } \sum_{k=1}^{\infty} ( \rho_{k} )^{2} < \infty	.
	\end{equation}
\end{assumption}

Specifically, the step size has to approach to zero when the time step approaches to infinity. With the policy gradient method, the algorithm will begin with an initial parameter vector $\Theta_{0} \in \mathfrak{R}^{|\mathcal{S}|}$, and the parameter vector $\Theta$ will be adjusted at each time step by using~(\ref{idealized_algorithm_theta}). With Assumption~\ref{derivatives} and Assumption~\ref{ass:step size}, as stated in~\cite{Bertsekas1995}, it is proved that $\lim_{k \rightarrow \infty} \nabla \xi(\Theta_{k}) = 0$, and thus $\xi(\Theta_{k})$ converges.

\subsection{Learning Algorithm}
By calculating the gradient of the function $\xi(\Theta_{k})$ with respect to $\Theta$ at each time step $k$, the average throughput $\xi(\Theta_{k})$ can be maximized based on the idealized gradient algorithm. Nevertheless, the gradient of the average throughput $\xi(\Theta_{k})$ may not be exactly calculated if the size of the state space $\mathcal{S}$ is very large. Therefore, the proposed online learning algorithm adopts an approach that can estimate the gradient $\xi(\Theta_{k})$ and update the parameter vector $\Theta$ at each time step as follows.

Under the constraint~(\ref{eq:condition_1}), with $\sum_{a \in \mathcal{A}} \chi_{\Theta}(s,a) = 1$, we have $\sum_{a \in \mathcal{A}} \nabla \chi_{\Theta}(s,a) = 0$. Hence, from~(\ref{eq:average_throguhput_defination}), $\nabla {\mathcal{T}}_{\Theta} (s)$ can be expressed as:
\begin{equation}
\begin{aligned}
\nabla {\mathcal{T}}_{\Theta} (s)  & =\sum_{a \in \mathcal{A}} \nabla \chi_{\Theta}(s, a) {\mathcal{T}} ( s, a)\\
&=\sum_{a \in \mathcal{A}} \nabla \chi_{\Theta}(s, a) ({\mathcal{T}} ( s, a )  - \xi(\Theta)).
\end{aligned}
\end{equation}

In addition, for all $s \in \mathcal{S}$, we have:
\begin{equation}
\begin{aligned}
&\sum_{s' \in \mathcal{S}}\nabla P_{\Theta}({s,s'}) d(a',\Theta) =\\
&\sum_{s' \in \mathcal{S}} \sum_{a \in \mathcal{A}} \nabla \chi_{\Theta}(s, a) P_{a}({s,s'}) d(s',\Theta).
\end{aligned}
\end{equation}


Then, under Proposition~\ref{prop_policy_gradient}, the gradient of $\xi(\Theta)$ can be expressed as follows:
\begin{equation}
\begin{aligned}
\nabla \xi(\Theta)  = &	\sum_{s \in \mathcal{S}} \pi_{\Theta}(s) \Big(\nabla {\mathcal{T}}_{\Theta} ( s ) + \sum_{s' \in \mathcal{S}}\nabla P_{\Theta}({s,s'}) d(s',\Theta) \Big) \\
= &	\sum_{s \in \mathcal{S}} \pi_{\Theta}(s)\Big(\sum_{a \in \mathcal{A}} \nabla \chi_{\Theta}(s, a) \big({\mathcal{T}} ( s, a ) - \xi(\Theta)\big)\\		
&+ \sum_{s' \in \mathcal{S}}\sum_{a \in \mathcal{A}} \nabla \chi_{\Theta}(s, a) P_{a}({s,s'}) d(s',\Theta) \Big) \\
= &	\sum_{s \in \mathcal{S}} \sum_{a \in \mathcal{A}} \pi_{\Theta}(s) \nabla \chi_{\Theta}(s, a) q_{\Theta}(s,a),
\end{aligned}
\end{equation}
where 
\begin{equation}
\begin{aligned}
q_{\Theta}(s,a)&= \Big({\mathcal{T}} ( s, a ) - \xi(\Theta)\Big) + \sum_{s' \in \mathcal{S}}P_{a}({s,s'}) d(s',\Theta)\\
&= \mathbb{E}_{\Theta} \Bigg[\sum_{k=0}^{T-1}\big( {\mathcal{T}} ( s_k, a_k) - \xi(\Theta) \big) | s_{0} = s, a_{0}=a \Bigg].
\end{aligned}
\end{equation}

Here $T=\min\{k>0 | s_{k}=s^{*}\}$ is the first future time that the learning algorithm visits the recurrent state $s^{*}$. In addition, $q_{\Theta}(s,a)$ can be expressed as the differential throughput if the ST chooses action $a$ at state $s$ based on policy $\chi_{\Theta}$. Then, we introduce Algorithm~\ref{algorithm0} that updates the parameter vector $\Theta$ at each time it visits the recurrent state $s^*$.
\begin{algorithm}
	\caption{Algorithm to update parameter vector $\Theta$ at each time it visits the recurrent state $s^*$}
	\label{algorithm0}
	\begin{algorithmic}[1]
		\State \textbf{Inputs:} $\nu$, $\rho_m$, and $\Theta_{0}$.
		\State \textbf{Initialize:} initiate parameter vector $\Theta_{0}$ and randomly select a policy for the ST.
		\For{\textit{k=1 to T}}
		\State{Update current state $s$}
		\If{$s_k \equiv s^*$}
		\begin{equation}
		\label{theta_al1}
		\Theta_{m+1} = \Theta_{m} + \rho_{m}F_{m}(\Theta_{m},\widetilde{\xi}_{m}),
		\end{equation}
		\begin{equation}
		\label{xi_al1}
		\widetilde{\xi}_{m+1} = \widetilde{\xi}_{m} + \nu \rho_{m}\sum_{k'=k_{m}}^{k_{m+1}-1}\Big({\mathcal{T}}(s_{k'}, a_{k'}) - \widetilde{\xi}_{m}\Big),
		\end{equation}
		\quad \quad where
		\begin{equation}
		F_{m}(\Theta_{m},\widetilde{\xi}_{m}) = \sum_{k'=k_{m}}^{k_{m+1}-1} \widetilde{q}_{\Theta_{m}}(s_{k'},a_{k'}) \frac{\nabla \chi_{\Theta_{m}}(s_{k'},a_{k'})}{\chi_{\Theta_{m}}(s_{k'},a_{k'})},
		\end{equation}
		\begin{equation}
		\widetilde{q}_{\Theta_{m}}(s_{k'},a_{k'}) = \sum_{k=k'}^{k_{m+1}-1}\Big({\mathcal{T}} (s_{k}, a_{k}) - \widetilde{\xi}_{m}\Big).
		\end{equation}
		\State{$m=m+1$}
		\EndIf
		\State{Update $\rho_m$}
		\EndFor
		\State {\textbf{Outputs:}} The optimal value of $\Theta$
	\end{algorithmic}
\end{algorithm}
%
In Algorithm~\ref{algorithm0}, the step size $\rho_{m}$ satisfies Assumption~\ref{ass:step size} and $\nu$ is a positive constant. The gradient of the randomized parameterized policy function in (\ref{eq:randomized_parameterized_function}) is derived as $\nabla \chi_{\Theta_{m}}(s_{k'},a_{k'})$. Additionally, $F_{m}(\Theta_{m},\widetilde{\xi}_{m})$ is the estimated gradient of the average throughput calculated by the cumulative sum of the total estimated gradient of the average throughput between the $m$-th and $(m+1)$-th visits of the algorithm to the recurrent state $s^*$. Through Algorithm~\ref{algorithm0}, the parameter vector $\Theta$ and the estimated average throughput $\widetilde{\xi}$ are adjusted iteratively. Then, the convergence result of Algorithm~\ref{algorithm0} is derived as in Proposition~\ref{prop2}.
\begin{proposition}
\label{prop2}
Under Assumption~\ref{recurrent_state}-\ref{ass:step size}, let $(\Theta_{0}, \Theta_{1}, \ldots, \Theta_{\infty})$ be a sequence of the parameter vectors generated by Algorithm~\ref{algorithm0}. Then, $\xi({\Theta_{m}})$ converges and
\begin{equation}
\lim_{m\rightarrow \infty} \nabla \xi(\Theta_{m}) = 0,
\end{equation}
with probability one. 
\end{proposition}
The proof of Proposition~\ref{prop2} can be found in~\cite{Marbach2001} and~\cite{Bertsekas1995}. Specifically, based on the stochastic approximation method~\cite{Borkar2008Stochastic}, it is proved that $\xi(\Theta)$ and $\widetilde{\xi}(\Theta)$ converge to a common limit. Then, the process of updating the parameter vector $\Theta$ can be expressed as a gradient method with diminishing errors, thereby we can prove that $\nabla \xi(\Theta_m)$ converges to $0$, i.e., $\nabla_{\Theta} \xi(\Theta_{\infty})=0$.

With Algorithm~\ref{algorithm0}, we need to store all values of $\frac{\nabla \chi_{\Theta_{m}}(s_{k},a_{k})}{\chi_{\Theta_{m}}(s_{k},a_{k})}$ and  $\widetilde{q}_{\Theta_{m}}(s_{k},a_{k})$ between the $m$-th and $(m+1)$-th visits in order to update the values of the parameter vector $\Theta$. This may lead to a slow processing especially when the size of the state space $\mathcal{S}$ is large. To deal with this shortcoming, the Algorithm~\ref{algorithm0} is modified to be able to update parameter vectors iteratively with simple calculations. First, we reformulate $F_{m}(\Theta_{m},\widetilde{\xi}_{m})$ as follows:
\begin{equation}
\begin{aligned}
&F_{m}(\Theta_{m},\widetilde{\xi}_{m})= \sum_{k'=k_{m}}^{k_{m+1}-1} \widetilde{q}_{\Theta_{m}}(s_{k'},a_{k'}) \frac{\nabla \chi_{\Theta_{m}}(s_{k'},a_{k'})}{\chi_{\Theta_{m}}(s_{k'},a_{k'})} \\
&= \sum_{k'=k_{m}}^{k_{m+1}-1}\frac{\nabla \chi_{\Theta_{m}}(s_{k'},a_{k'})}{\chi_{\Theta_{m}}(s_{k'},a_{k'})} \sum_{k=k'}^{k_{m+1}-1}\big( {\mathcal{T}} (s_{k}, a_{k}) - \widetilde{\xi}_{m}\big)\\
&=\sum_{k'=k_{m}}^{k_{m+1}-1} \big( {\mathcal{T}} (s_{k}, a_{k}) - \widetilde{\xi}_{m}\big) z_{k+1}, 
\end{aligned}
\end{equation}
where
\begin{equation}
z_{k+1} = \left\{ 
\begin{array}{ll}
\frac{\nabla \chi_{\Theta_{m}}(s_{k},a_{k})}{\chi_{\Theta_{m}}(s_{k},a_{k})}, & \text{if} \phantom{1} k = k_{m},\\
z_{k}+\frac{\nabla \chi_{\Theta_{m}}(s_{k},a_{k})}{\chi_{\Theta_{m}}(s_{k},a_{k})}, & k=k_{m}+1,\ldots,k_{m+1}-1. \\
\end{array}
\right.
\label{eq:zupdate_time}
\end{equation}

Then, the algorithm now can be expressed as in Algorithm~\ref{algorithm1}, where $\nu$ is a positive constant and $\rho_{k}$ is the step size of the algorithm. 
\begin{algorithm}
	\caption{Algorithm to update $\Theta$ at every time step}
	\label{algorithm1}
	\begin{algorithmic}[1]
		\State \textbf{Inputs:} $\nu$, $\rho_k$, and $\Theta_{0}$.
		\State \textbf{Initialize:} initiate parameter vector $\Theta_{0}$ and randomly select a initial policy for the ST.
		\For{\textit{k=1 to T}}
		\State{Update current state $s_k$}
		\State{}
		\begin{equation}
		z_{k+1} = \left\{ 
		\begin{array}{ll}
		\frac{\nabla \chi_{\Theta_{k}}(s_{k},a_{k})}{\chi_{\Theta_{k}}(s_{k},a_{k})}, & \text{if} \phantom{1} s_{k} = s^{*},\\
		z_{k}+\frac{\nabla \chi_{\Theta_{k}}(s_{k},a_{k})}{\chi_{\Theta_{k}}(s_{k},a_{k})}, & \text{otherwise,} \\
		\end{array}
		\right.
		\label{eq:zupdate_state}
		\end{equation}
		\begin{equation}
		\Theta_{k+1} = \Theta_{k} + \rho_{k}({\mathcal{T}}( s_k, a_k ) -\widetilde{\xi}_{k})z_{k+1}, 
		\label{eq:updatetheta}
		\end{equation}
		\begin{equation}
		\widetilde{\xi}_{k+1} = \widetilde{\xi}_{k} + \nu\rho_{k}({\mathcal{T}}( s_k, a_k ) - \widetilde{\xi}_{k}).
		\label{eq:updatexi}
		\end{equation}
		\State{Update $\rho_k$}
		\EndFor
		\State {\textbf{Outputs:}} The optimal value of $\Theta$
		
	\end{algorithmic}
\end{algorithm}
Instead of calculating the value of $\frac{\nabla \chi_{\Theta_{k}}(s_{k},a_{k})}{\chi_{\Theta_{k}}(s_{k},a_{k})}$ directly, we can use some mathematical manipulation to transform it into an equivalent form by $1-\chi_{\Theta}(s,a)$. Thus, at each computing step, the ST just needs to perform basic calculations without any complex functions, thereby the online learning algorithm can be efficiently implemented on power-constrained devices.

\section{Performance Evaluation} 



\subsection{Experiment Setup}
We perform the simulations using MATLAB to evaluate the performance of the proposed solution under different parameter settings. In particular, when the primary channel is busy, we assume that if the secondary transmitter (ST) harvests energy, it can successfully harvest one unit of energy with probability 0.9. Otherwise, if the ST performs backscattering to transmit data, it can successfully transmit one unit of data with probability 0.9. When the channel is idle and if the ST wants to transmit data actively, the ST requires one unit of energy to transmit two units of data. The successful data transmission probability when the channel is idle is also assumed to be 0.9. The maximum data size and the energy storage capacity are set to be 10 units. Unless otherwise stated, the idle channel probability and the packet arrival probability are 0.5. For the learning algorithm, i.e., Algorithm~\ref{algorithm1}, we use the following parameters for the performance evaluation. At the beginning, the ST will start with a randomized policy, i.e., stay idle or transmit data if the primary channel is idle, and harvest energy or backscatter data otherwise. We set the initial value of $\rho=0.00001$ and it will be updated after every 18,000 iterations as follows: $\rho_{k+1} = 0.9 \rho_{k}$. We also set $\nu = 0.01$. To evaluate the proposed solution, we compare its performance with three other schemes, i.e., optimal policy~\cite{Puterman1995}, HTT policy~\cite{park2013}, and backscatter policy~\cite{LiuAmbient2013}. The optimal policy is obtained through using the value iteration algorithm when all environment information is available in advance. The optimal policy will be used as a benchmark to evaluate the performance of the proposed learning algorithm when the environment information is not available in advance. 

\subsection{Numerical Results}
\subsubsection{Convergence of the learning algorithm}
\begin{figure}[h]
	\centering
	\includegraphics[scale=0.25]{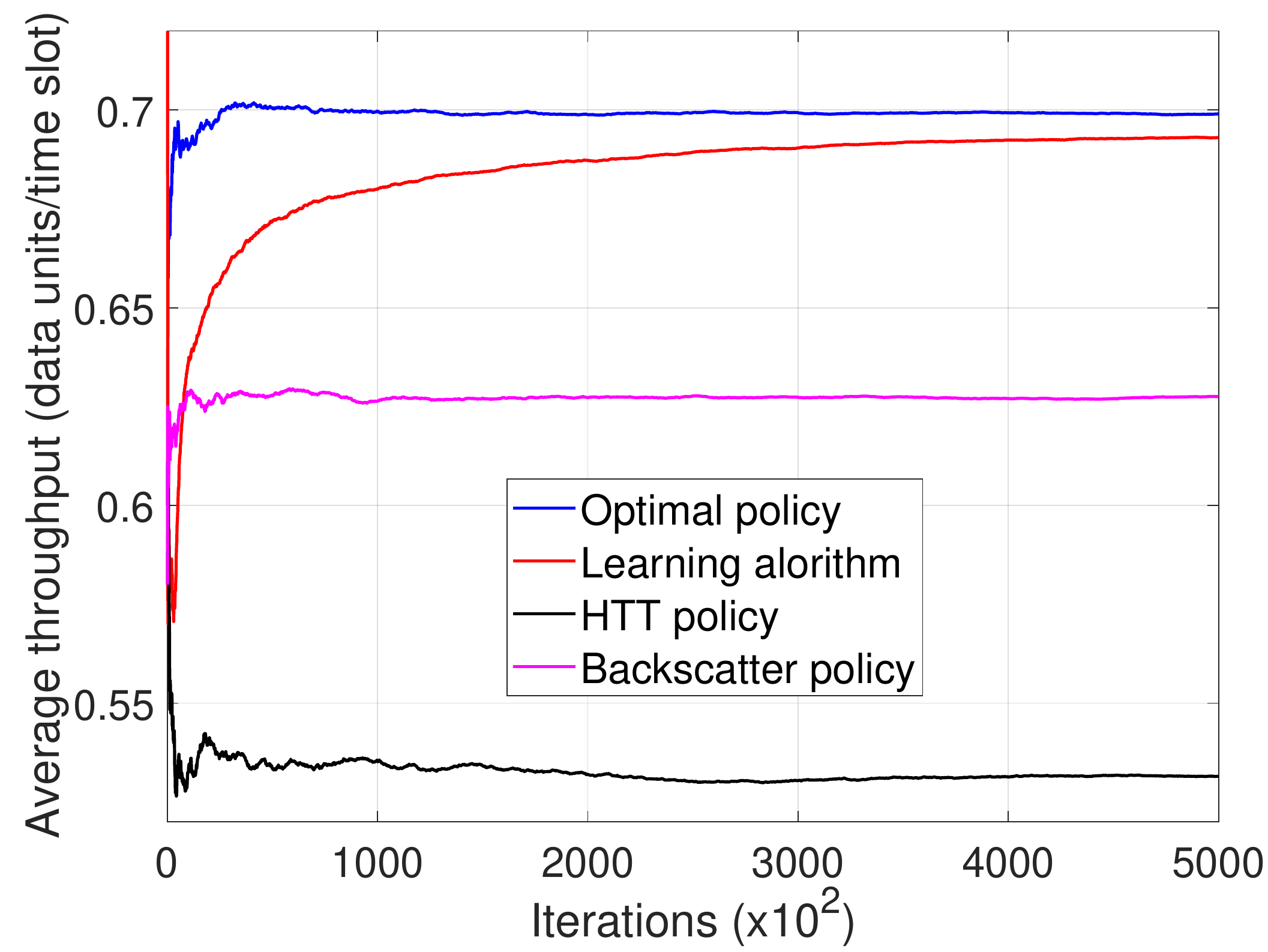}
	\caption{The convergence of the learning algorithm.} 
	\label{fig:convergence}
\end{figure}

We first show the learning process and the convergence of the proposed algorithm. As shown in Fig.~\ref{fig:convergence}, the performance of the ST is fluctuated in the first 4,000 iterations as the ST is still learning to adjust the parameter $\Theta$. After that, the learning process begins to stabilize, and then the average throughput converges to $0.68$ after $10^5$ iterations. 

\begin{figure}[h]
	\centering
	\begin{subfigure}[b]{0.3\textwidth}
		\centering
		\includegraphics[scale= 0.25]{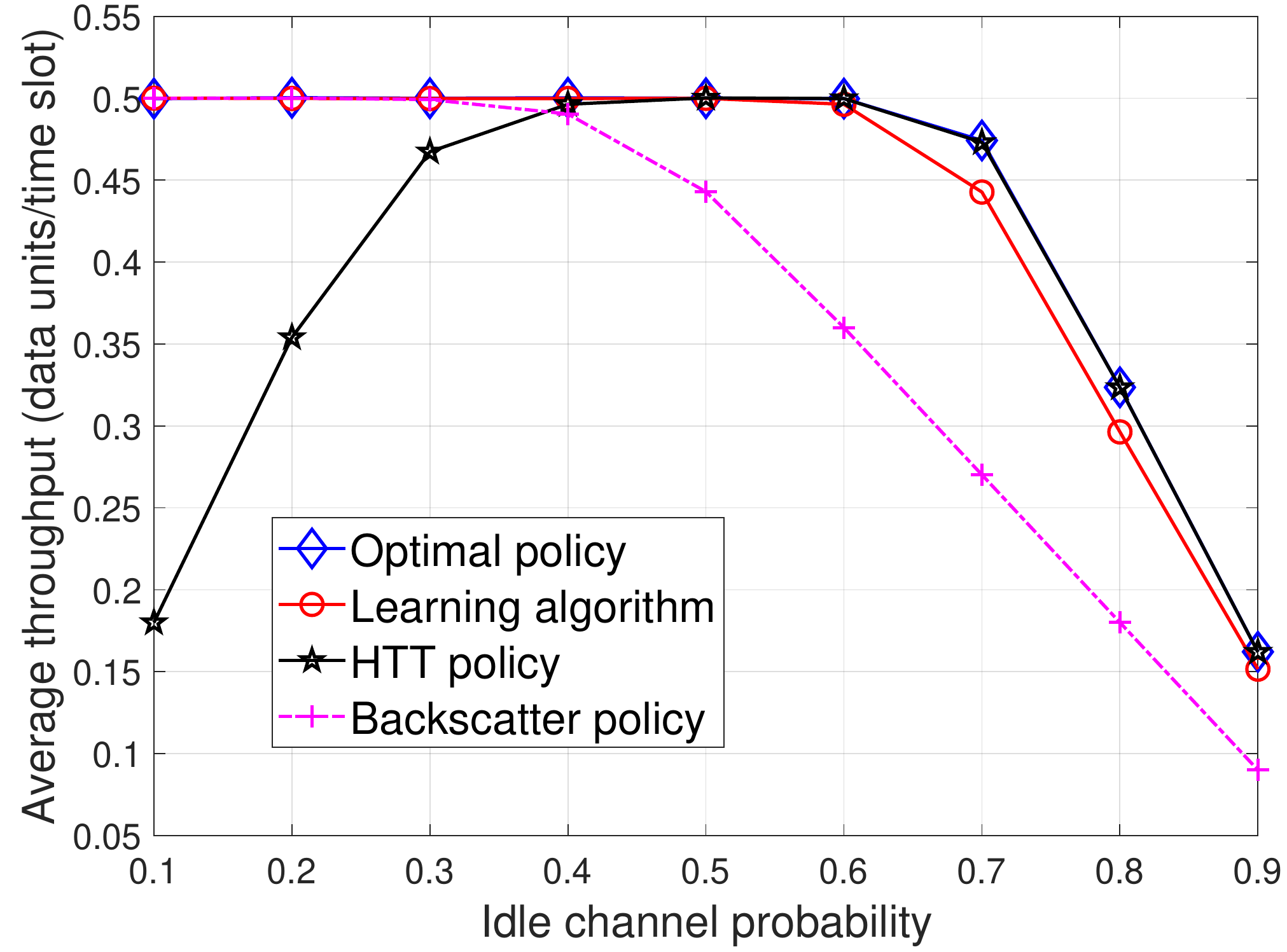}
		\caption{Idle channel probability is varied}
	\end{subfigure}%

	\begin{subfigure}[b]{0.3\textwidth}
		\centering
		\includegraphics[scale= 0.25]{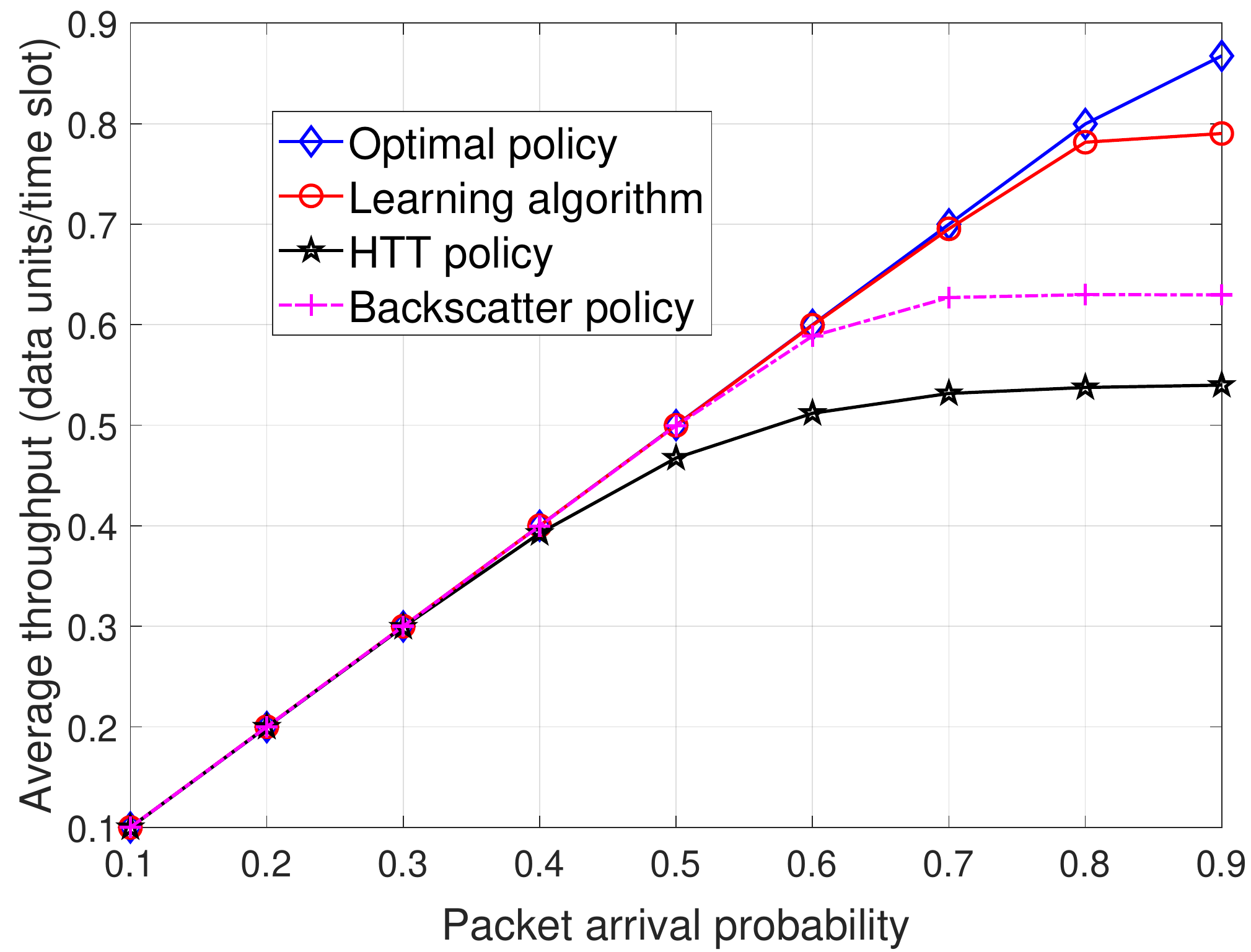}
		\caption{Packet arrival probability is varied}
	\end{subfigure}
	\caption{The average throughput of the ST.} 
	\label{fig:throughput}
\end{figure}
\subsubsection{Network performance}
\begin{figure}[h]
	\centering
	\begin{subfigure}[b]{0.3\textwidth}
		\centering
		\includegraphics[scale= 0.25]{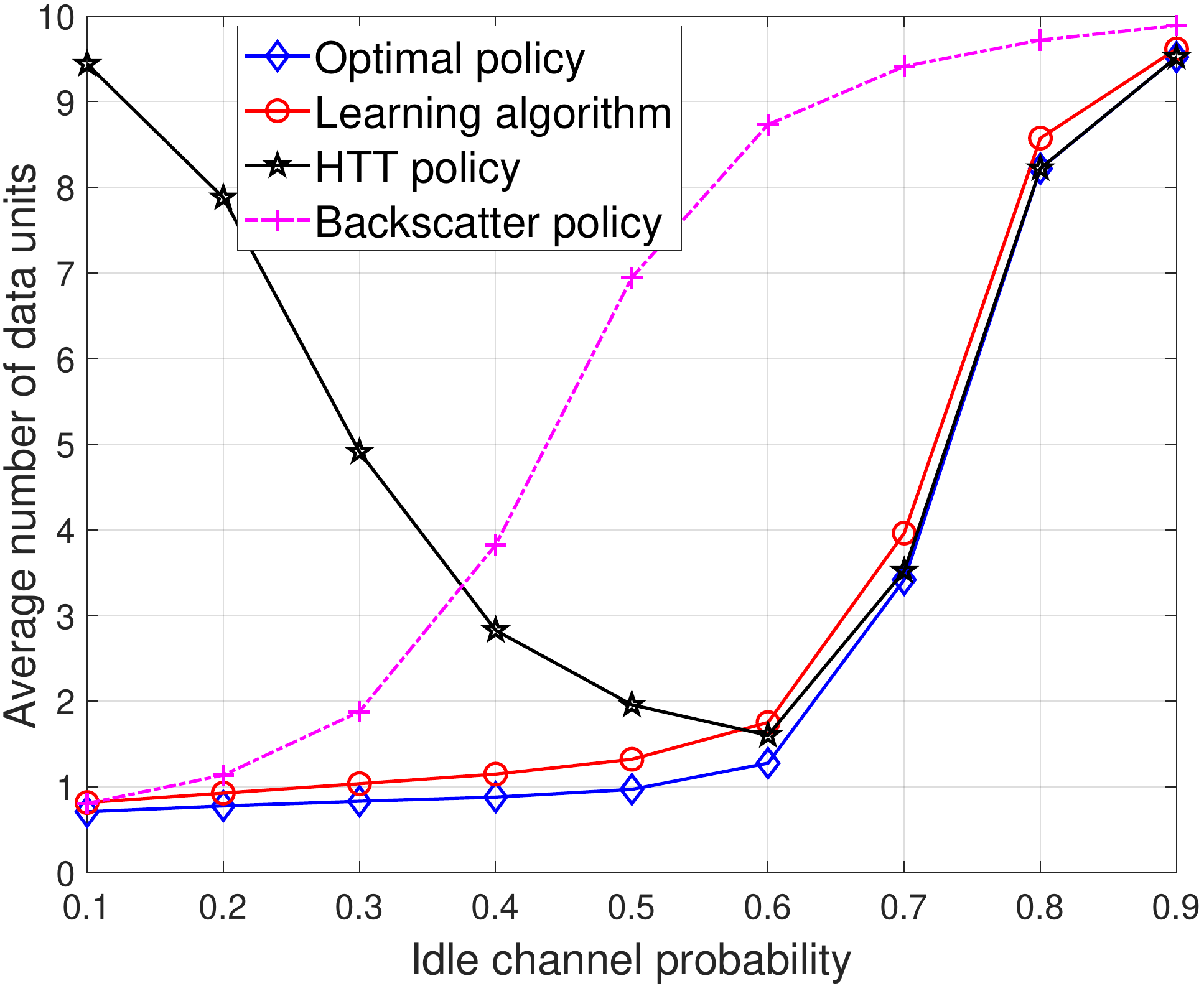}
		\caption{}
	\end{subfigure}%
	
	\begin{subfigure}[b]{0.3\textwidth}
		\centering
		\includegraphics[scale= 0.25]{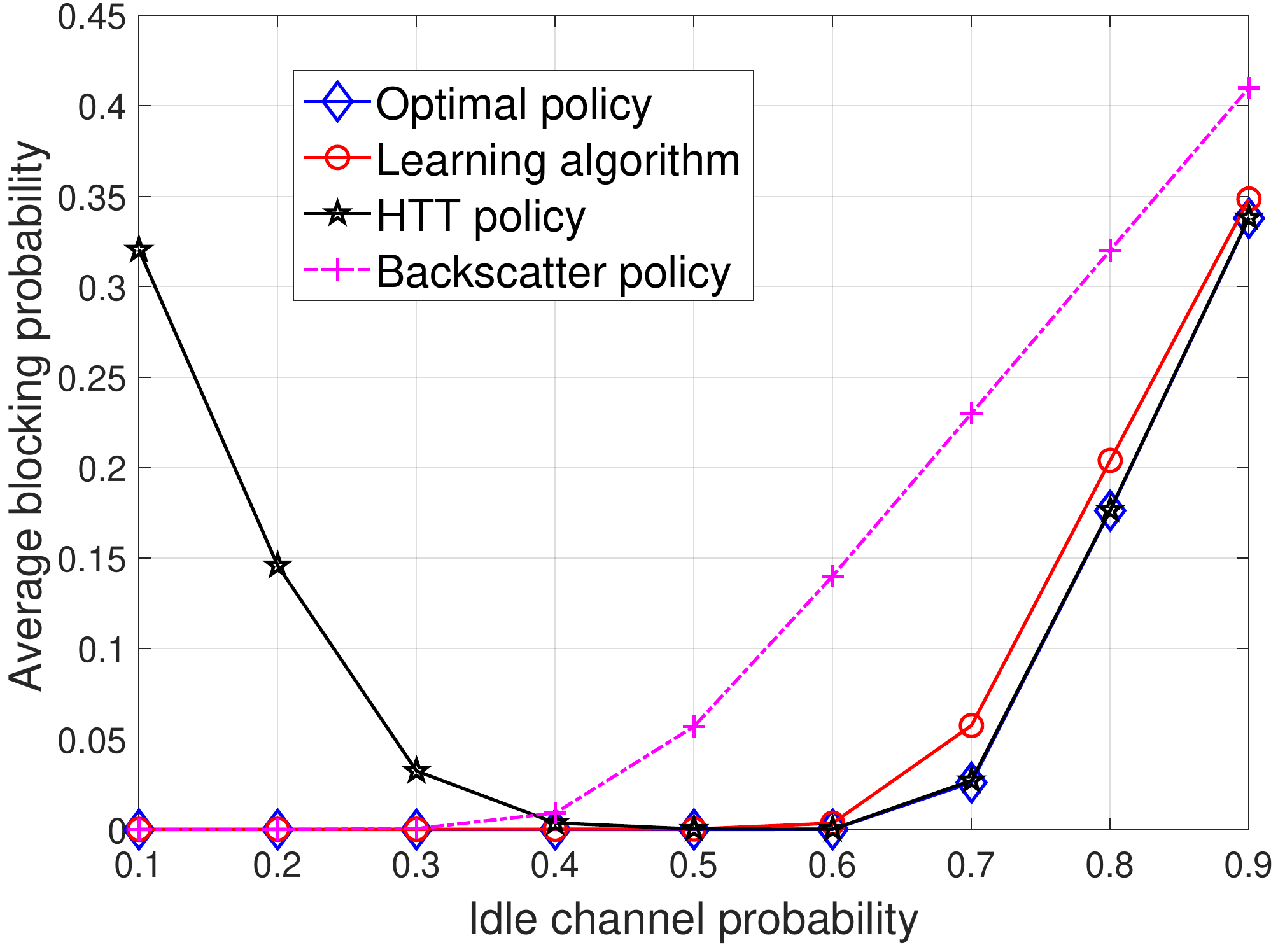}
		\caption{}
	\end{subfigure}%
	\caption{(a) The average number of data units in the data queue and (b) blocking probability.} 
	\label{fig:packetblocking}
\end{figure}
Next, we perform simulations to evaluate the performance of the proposed solution, i.e., Algorithm~\ref{algorithm1}, and compare with the three other policies, i.e., the optimal, HTT, and backscatter policies, in terms of the average throughput, delay, and blocking probability. In Figs.~\ref{fig:throughput}(a) and~\ref{fig:throughput}(b), we show the average throughput of the ST obtained by different policies when the idle channel and packet arrival probabilities are varied. Obviously, when the channel idle probability increases, the average throughput of the ST decreases accordingly. However, the learning algorithm always achieves the throughput close to that of the optimal policy. Note that, when the idle channel probability is low, i.e., less than $0.5$, the average throughput obtained by HTT policy increases. This is from the fact that the ST has higher opportunities to transmit data as the primary channel is likely to be idle. Nonetheless, when the idle channel probability is high, i.e., higher than $0.6$, the throughput obtained by HTT policy decreases as the ST has little time to harvest energy for data transmission process. Similarly, in Fig.~\ref{fig:throughput}(b), the throughputs of all the policies increase when the packet arrival probability increases. When the packet arrival probability is higher than 0.4, the optimal policy achieves the highest throughput followed by the learning algorithm.

We then investigate the blocking probability and delay of all policies as shown in Fig.~\ref{fig:packetblocking}. Clearly, when the idle channel probability increases, the average number of data units in the data queue and the blocking probability also increase. This is due to the fact that the ST has less opportunities to backscatter data and does not have sufficient energy to transmit data to its receiver as the primary channel is likely to be idle. However, the proposed learning algorithm always achieves the performance close to that of the optimal policy.

\section{Summary} 
In this paper, we have considered the RF-powered backscatter cognitive radio network in which the secondary transmitter is equipped with wireless energy harvesting and backscattering capabilities. In this network, the secondary transmitter can harvest energy or backscatter data to its receiver when the channel is busy. To maximize the network performance, we propose an online learning algorithm that enables the secondary transmitter to adjust its decision to obtain the optimal policy by interacting with the environment. Through numerical results, we have demonstrated that the proposed solution can achieve performance better than the conventional methods and close to that of the optimal policy without requiring the complete information from the environment in advance.


\section*{Acknowledgment}
This work was supported in part by WASP/NTU M4082187 (4080), Singapore MOE Tier 1 under Grant 2017-T1-002-007 RG122/17, MOE Tier 2 under Grant MOE2014-T2-2-015 ARC4/15, NRF2015-NRF-ISF001-2277, and EMA Energy Resilience under Grant NRF2017EWT-EP003-041.

\bibliographystyle{IEEE}

\end{document}